\documentclass{aa}
\usepackage{color}
\usepackage{graphicx}
\usepackage{txfonts}
\usepackage{natbib}
\bibpunct{(}{)}{;}{a}{}{,}

\newcommand{\IR}{IRC\,+10\,420~}
\newcommand{\Brg}{Br$\gamma$~}

\def\mathstacksym#1#2#3#4#5{\def#1{\mathrel{\hbox to 0pt{\lower#5\hbox{#3}\hss} \raise #4\hbox{#2}}}}
\mathstacksym\gta{$>$}{$\sim$}{1.5pt}{3.5pt} 
\mathstacksym\lta{$<$}{$\sim$}{1.5pt}{3.5pt} 
\begin{document}
                                %

\title{Neutral and ionized  gas around the post-Red Supergiant IRC~+10\,420  at au size scales\thanks{Based on
observations at ESO, and in particular with VLTI, proposals 079.D-0123(A), and
 383.C-0166(A) and X-Shooter, proposal SV-9434}}

\titlerunning{AMBER observations of the massive evolved star IRC\,+10\,420}
\author{ Ren\'e D. Oudmaijer\inst{1}, W.J. de Wit\inst{2}} 
\offprints{R.D. Oudmaijer, \email{r.d.oudmaijer@leeds.ac.uk}}
\institute{School of Physics \& Astronomy, University of Leeds, Woodhouse Lane,
                                Leeds LS2 9JT, UK\and
               European Southern Observatory, Casilla 19001, Santiago 19, Chile            
}          
        
\date{Received date; accepted date}
\abstract
{\IR is one of the few known massive stars in rapid transition from
  the Red Supergiant phase to the Wolf-Rayet or Luminous Blue Variable
  phase.}
{The star has an ionised wind and using the \Brg line we assess the
mass-loss on spatial scales of $\sim 1$\,au.}
{We present new VLT Interferometer AMBER data which are combined with
  all other AMBER data present in the literature.  The final dataset
  covers a position angle range of $\sim 180\degr$ and baselines up to
  110 meters.  The spectrally dispersed visibilities, differential
  phases and line flux are conjointly analyzed and modelled.  We also
  present the first AMBER/FINITO observations which cover a larger
  wavelength range and allow us to observe the \ion{Na}{i} doublet at
  2.2 $\mu$m. The data are complemented by X-Shooter data, which
  provide a higher spectral resolution view.}
{ The \Brg emission line and the \ion{Na}{i} doublet are
  both spatially resolved. After correcting the AMBER data
  for the fact that the lines are not spectrally resolved, we find
  that \Brg  traces a ring  with a diameter of
  4.18 mas, in agreement with higher spectral resolution data. We
  consider a geometric model in which the \Brg emission emerges from
  the top and bottom rings of an hour-glass shaped structure, viewed
  almost pole-on. It provides satisfactory fits to most visibilities
  and differential phases. The fact that we detect line emission from
  a neutral metal like \ion{Na}{i}  within the ionized region, a
  very unusual occurrence, suggests the presence of a dense
  pseudo-photosphere.  }
{The ionized wind can be reproduced with a polar wind, which could
well have the shape of an hour-glass. Closer in, the resolved
\ion{Na}{i} emission is found to occur on scales barely larger than
the continuum. This fact and that many Yellow Hypergiants
exhibit this comparatively rare emission hints at the presence of a
``Yellow'' or even ``White Wall'' in the Hertzsprung-Russell diagram, preventing them
from visibly evolving to the blue.}
\keywords{stars: evolution - stars: mass loss - supergiants - stars: individual IRC
  +10\,420 - techniques: interferometric} 
\maketitle
\section{Introduction}
\label{intro}

The star \IR is of central importance to our understanding of the
massive star evolution across the Hertzsprung-Russell diagram 
from the Red Supergiant to the Wolf-Rayet or Luminous Blue Variable
(LBV) phase. There are strong indications that its photospheric
temperature increased very rapidly, by $\sim$2200\,K over 30
years. \IR has currently attained mid-A spectral type (see Oudmaijer
1998; Klochkova et al. 2002\nocite{2002ARep...46..139K}).  Its
distance of 3.5 to 5\,kpc implies a luminosity typical for a star with
an initial mass of around $\rm 40\,M_{\odot}$, which places it close
to the Humphreys-Davidson limit in the Hertzsprung-Russell diagram (Jones et
al. 1993\nocite{1993ApJ...411..323J}).  The spectral type makes \IR a
member of the class of Yellow Hypergiants \citep{dejager}, while its
infrared excess due to circumstellar dust indicates that it was only
recently undergoing extreme mass-loss, presumably in the Red
Supergiant phase (Oudmaijer et
al. 1996\nocite{1996MNRAS.280.1062O}). The number of known post-Red
Supergiants is very small, Oudmaijer et
al. (2009)\nocite{2009ASPC..412...17O} discussed two such examples,
\IR and HD 179821, while \citet{lagadec} recently proposed that IRAS
17163-3907 (the ``Fried Egg'' nebula, so named because of its double
shell structure) belongs to this class too.

The intricate circumstellar environment of \IR has various distinctive
components on different spatial scales, an overview of which can be
found in de Wit et al. (2008\nocite{2008A&A...480..149D}), see also
Tiffany et al. (2010\nocite{2010AJ....140..339T}). Of importance to
the study presented here is the fact that the dusty environment
responsible for the strong near-infrared excess has been spatially
resolved with speckle interferometry at 2.11$\mu$m (Bl\"{o}cker et
al. 1999\nocite{1999A&A...348..805B}). They found the inner radius of
the dust shell to be 70\,milli-arcsecond (mas), consistent with the
dust condensation temperature. In addition, they found that about 60\%
of the $K$-band continuum emission remains unresolved (see also
Monnier et al. 2004\nocite{2004ApJ...605..436M}), and this emission is
identified with the photospheric radiation from the star
itself. Further 1D-modelling led them to conclude that \IR is
surrounded by two separate shells centred on the star with diameters
of $0.070\arcsec$ and $0.310\arcsec$.

Interior to the dusty environment, \IR displays an ionised wind, which
is unusual for its spectral type. Two dedicated studies using the VLT
Interferometer (VLTI) and the near-infrared AMBER instrument have
aimed at elucidating the issue of the shape of the ionised wind on
milli-arcsecond spatial scales.  De Wit et al. (2008) show that the
\Brg emission region is spatially resolved and derive a size of
3.3\,mas along the employed VLTI baseline. Under the assumption that
the transition is optically thick, they find that the total observed
line flux is too small if the emitting region were a circular disk on
the sky. Instead, de Wit et al. conjecture that it is possibly
elongated, a geometry with the same radius, but smaller projected
area, analogous to the H$\alpha$ emission (Davies et al. 2007).
Alternatively, the total line flux can be accounted for if the
geometry resembles a ring-like structure, which is proposed in Driebe
et al. (2009\nocite{2009A&A...507..301D}, hereafter D09). The latter
publication presents observations employing two triplet VLTI
configurations using the Auxiliary Telescopes for a single position
angle on the sky and with AMBER in low spectral resolution mode. A
stellar diameter of approximately 1.0 milli-arcsecond was determined
using this dataset. These authors also present a single Unit Telescope
(UT) triplet with AMBER in high spectral resolution, resolving both
spatially and spectrally the \Brg emission line of IRC\,+10\,420. The
differential phase of the spectrally resolved line (probing the
photo-centre) shows an intriguing profile and the authors' efforts to
reproduce the observables with two-dimensional radiative transfer
modelling remained unsuccessful. As a corollary, the geometry of the
ionized wind and mass-loss of \IR on milli-arcsecond scales continues
to pose a serious challenge.

\begin{figure}
 \includegraphics[height=8cm,width=8cm]{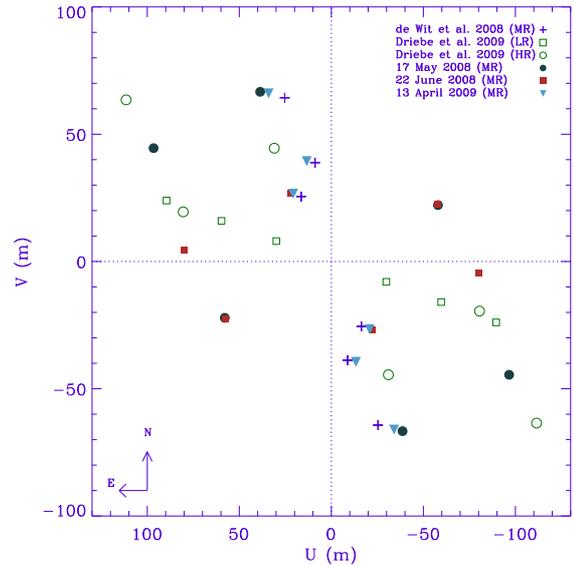}
 \caption[]{{\it uv}-plane coverage of VLTI/AMBER observations of IRC\,+10\,420. Shown
are the low and high spectral resolution observations by D09 and 
the medium resolution observations presented in de Wit et al. (2008) and in this 
paper. }
 \label{uvplane}
\end{figure}

In this third VLTI/AMBER study of \IR we present six new baselines
which, together with the previously presented ones, add to a total
position angle coverage of $\sim 180\degr$ for the \Brg emission (see
Fig.\,\ref{uvplane}).  Our analysis of the ionized emission, which we
present here, includes all the previously published interferometric
AMBER data of the object.  In addition, the use of the fringe-tracker
FINITO on one set of baselines provides visibility and phase
information for a much larger wavelength range.  We report and
interpret for the first time spatial information of a neutral metallic
line in addition to the hydrogen recombination line.  The inclusion of
the high spectral resolution data in D09 allows us to re-assess the
much larger set of medium spectral resolution data in hand and we
therefore revisit the object and the data taken thus far. In addition,
we obtained X-Shooter data of IRC +10420 during the science
verification phase, and present the part of the spectrum that overlaps
with the AMBER data.

In Sect.\,\ref{observ} we present the new observations with the VLTI
and the data reduction of AMBER in medium spectral resolution mode. We
discuss in this section the effect of spectral resolution in case of
unresolved spectral lines on the derived characteristic size
scales. In Sect.\,\ref{results} we present the new results. In the
discussion section \ref{disc}, a geometrical model is presented. We
discuss the Na {\sc i} emission, and explain this in the framework of
the presence of a pseudo-photosphere. By extrapolation, the presence
of this, comparatively rare emission line, in the spectra of other
Yellow Hypergiants leads us to speculate that this is a general
feature, and that all such objects have a pseudo-photosphere.

\begin{table*}
  {
    \begin{center}
      \caption[]{Technical overview of the AMBER observations of
\IR. }
      \begin{tabular}{cccrrccccccc}
        \hline
        \hline
	
Config.      & 	Date      & Station & B & P.A.  & DIT & $\tau_{\rm coh}$ & $\rm <Seeing>$ & $\rm N_{files}$ & $\rm N_{frames}$& $V_{\rm cal}$ & $V_{Br\gamma}/V_{cont}$ \\ 
              &            &        & (m)  & ($^{\rm o})$ & (s) & (ms) & (arcsec) & \\            
\hline   
A            & 2008-05-17 & U1-U3   &  77.1    & 30.1 & 0.05              &  2.8            & 0.68    & 9      & 1000    & $0.85\pm0.07$   &   $0.75\pm0.01$   \\ 
             &            & U3-U4   &  62.0    & 110.9 &                  &                 &         &        &         & $0.88\pm0.07$   &   $0.79\pm0.01$   \\ 
             &            & U4-U1   & 106.3    & 65.3 &                   &                 &         &        &         & $0.86\pm0.07$   &   $0.58\pm0.005$   \\
\hline 
B            & 2008-06-22 & U2-U3   &  34.9    & 39.6 & 0.05              &  2.4            & 0.78    & 9      & 1000    & -    &  $0.93\pm0.005$   \\ 
             &            & U3-U4   &  61.8    & 111.2 &                  &                 &         &        &         & -    &  $0.82\pm0.01$    \\ 
             &            & U4-U2   &  80.0    &  86.8 &                  &                 &         &        &         & -    &  $0.70\pm0.01$   \\ 
\hline
C$^{1}$      & 2009-04-13 & U1-U2   &  41.8    & 19.2  & 0.187            &  4.2            & 1.1     & 5      & 200     &  $0.86\pm0.10$ &  $0.90\pm0.005$   \\ 
             &            & U2-U3   &  34.0    & 37.3  &                  &                 &         &        &         &  $0.97\pm0.07$ &    -              \\ 
             &            & U3-U1   &  74.8    & 27.8  &                  &                 &         &        &         &  $0.77\pm0.10$ &  $0.80\pm0.005$   \\ 
        \hline 
        \hline
      \end{tabular}
      \label{tabvs}
    \end{center}

Calibrated visibilities correspond to 20\% frame selection for
config.\,A, and 5\% for config.\,C. The Detector Integration Times
(DIT) are the integration times per individual exposure.
$^{1}$ Observations with FINITO.
  }
\label{obs}
\end{table*}

\section{Observations and data reduction}
\label{observ}

\subsection{AMBER spectro-interferometry}

New observations of \IR were obtained with the UTs and the AMBER
instrument (Petrov et al. 2007\nocite{2007A&A...464....1P}) on three
separate occasions.  All data were taken in the Medium Resolution (MR)
mode, while we present the first MR observations of the object taken
with the FINITO fringe-tracker (Gai et
al. 2004\nocite{2004SPIE.5491..528G}). The use of FINITO in this mode
extends the observed wavelength range substantially compared to the
previous studies, and includes the Na {\sc i} 2.2 $\mu$m doublet which
has long been known to be in emission (Thompson \& Boroson, 1977, and
see below). 

We observed \IR twice in service mode, on the nights of 17$^{th}$ May
and 22$^{nd}$ June 2008 (ESO cycle P79), and once in visitor mode on
the night of the of 13$^{th}$ April 2009 (ESO cycle P83). The VLTI
configurations on these occasions were the U1-U3-U4, U2-U3-U4 and
U1-U2-U3, respectively, which we further refer to as A, B, and C.  The
technical overview of the observations is given in
Table\,\ref{obs}. The {\it uv}-plane coverage of all the published
VLTI-AMBER data with various spectral resolutions taken to date is
shown in Fig.\,\ref{uvplane}. This includes the medium spectral
resolution observations published in de Wit et al. (2008) and the low
and high spectral resolution observations presented in D09.  Given its
Northern declination, this is arguably the most complete coverage
possible for the object.

The AMBER instrument was set-up in the medium spectral resolution mode
which provides a spectral resolution of 1500 or $\rm 200\,km\,s^{-1}$.
Configurations A and B cover a wavelength range between 2.12 and
2.19\,$\mu$m, and includes the Br$\gamma$ hydrogen recombination line
at 2.167\,$\mu$m. This transition was the main subject of the two
previous AMBER studies on \IR. The instrumental set-up of
configuration C was different as it included the fringe-tracker
FINITO. It allows higher precision interferometry, and a much larger
wavelength range can be probed and read-out. In the present case it
runs from basically 2.0 to 2.3\,$\mu$m.  This set-up therefore allows
the spatial study of the \ion{Na}{i} doublet at 2.206 and 2.209
$\mu$m, which is known to be in emission. The doublet was first
reported by Thompson \&  Boroson (1977\nocite{1977ApJ...216L..75T}),
and later by Hanson et al. (1996\nocite{1996ApJS..107..281H}) and
Humphreys et al. (2002\nocite{2002AJ....124.1026H}). Importantly, the
Na {\sc i} emission had already been detected before the onset of
Br$\gamma$ emission was discovered by Oudmaijer et
al. (1994\nocite{1994A&A...281L..33O}).  As the rest of the
AMBER-FINITO spectrum is featureless, we restrict ourselves in the
discussion to the 2.16-2.21 $\mu$m wavelength range which covers the
emission lines present.

All data were reduced using the {\it amdlib} (version 3.0) software
package dedicated to AMBER data and provided by Jean-Marie Mariotti
Centre.  Atmospheric jitter causes the interference fringe pattern to
be smeared out, leading to artificially low visibilities. Short
integration times are thus mandatory in order to limit smearing in the
absence of a fringe-tracker, and good atmospheric conditions
(coherence time, seeing) are essential.  A measure of successfully
interfering the three telescope beams and the recording of fringes is
provided by the {\it amdlib} package parameter signal-to-noise ratio
(SNR, Tatulli et al. 2007b\nocite{2007A&A...464...29T}). It is based
on the weighted summing of the correlated flux over all spectral
channels per integration time (a so-called interferogram). Currently,
the recommended approach regarding these data quality degrading
effects is to make a frame selection based on an SNR criterion, and
discarding low SNR frames up to a certain cut-off (e.g see Tatulli et
al. 2007a\nocite{2007A&A...464...55T}).  Overall the values of the
instrumental visibilities depend strongly on the fraction of frames
selected, as the visibilities are biased towards smaller values when
more frames are included. However, significant for a study such as the
present one, is that the differential visibilities, i.e. the
visibility of the emission lines compared to the continuum, are not
very sensitive to frame selection (see de Wit et al. 2008).

\begin{figure}
 \includegraphics[height=9cm,width=7cm,angle=90]{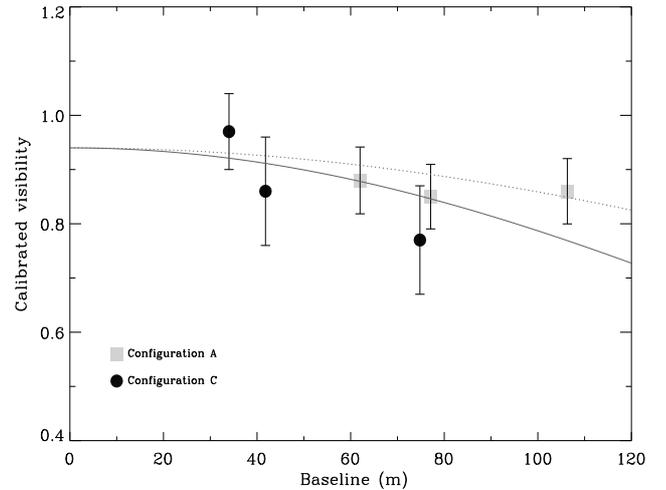}
 \caption[]{Calibrated continuum visibilities and their uncertainties
compared to a Gaussian shaped emitting region with FWHM of 0.98\,mas
(full line) and with FWHM of 0.7\,mas (dotted line). Both geometrical
models have a 6\% flux contribution to the continuum by an
over-resolved source.}
 \label{viscal}
\end{figure}

The observations in configuration A suffered from an abrupt change in
weather conditions. The coherence time $\tau_{\rm coh}$ nearly halved
in the course of the target observations, and the presence of thin
cirrus was reported in the meteorological report of these service
observations. Assessment of the data quality led us to consider only
the first four files of a set of in total nine observation files each
consisting of 1000 interferograms. The $<\tau_{\rm coh}>$ in Table\,1
refers to these four files. The equally bright calibrator star
\object{HD\,175743} ($K=3.5^{m}$,  uniform disk diameter 1.012 mas;
\citealt{lafrasse}) was observed under relatively constant weather
conditions, albeit with a somewhat smaller coherence time than during
the target observations. The calibrator star has a K\,1\ giant
spectral type, and does not show any spectral absorption features in
its flux spectrum at the employed spectral resolution. It was observed
within 30 minutes of the target and again nine files were
secured. From the spread around the mean, we determined a statistical
uncertainty of 4-5\% on the squared visibilities for any frame
selection criterion. The error on the calibrated visibilities is
7\%. The calibrated continuum visibilities for configurations A and C
are presented in Fig.\,\ref{viscal}. They are nearly constant as
function of baseline length, and this would indicate that the size of
the continuum is somewhat smaller than the one proposed in D09.

The target observations in configuration B were performed under stable
atmospheric conditions, although some thin cirrus was present. Again
nine observations were performed with similar data quality. We retain
all files in our analysis. The same calibrator star as for
configuration A, \object{HD\,175743}, was used, but only five
observation sets were secured. The weather conditions were relatively
stable and the calibrator observations were performed within 15
minutes of the target object. A statistical uncertainty between
4.5-5.5\% on the squared visibilities of the calibrator was determined
from the frame selection criterion. However, the absolute calibration
of the visibilities delivered visibilities in excess of unity for any
applied frame selection (for example they were of order 1.1 for a 20\%
frame selection) and are, obviously, unphysical. The instrument set-up
and observing logs were checked, but no obvious reasons for this
problem could be identified. We discard the absolute calibration for
this set of observations.

Observations performed in configuration C were done with the
assistance of the fringe tracker FINITO (Gai et
al. 2004\nocite{2004SPIE.5491..528G}), The use of the fringe tracker
stabilizes the position of the fringes on the AMBER detector allowing
a longer integration time and a better SNR.  The coherence time was
stable for the observations, but the seeing was rather poor
between 1.05 and 1.25\arcsec. As a result, FINITO was not able to lock
continuously on the fringes, and, given the longer integration times
on AMBER, the frame selection criterion had to be chosen more
stringent. We retained only 5\% of the frames leading to quite 
large uncertainties in the observables as shown in
Fig.\,\ref{vis3}. The baseline configuration of this dataset is close
to the one published in de Wit et
al. (2008\nocite{2008A&A...480..149D}).  In total 5 files of 200
frames were secured for both \IR and the somewhat brighter G8\,III
calibrator \object{HD\,194013} (K=$3.0^{m}$, uniform disk diameter 1.152 mas;
\citealt{lafrasse}). Stars of the latter
spectral type and luminosity class show \Brg and \ion{Na}{i}
absorption, albeit quite minor (Ivanov et
al. 2004\nocite{2004ApJS..151..387I}).

In Fig.\,\ref{viscal} the calibrated AMBER continuum visibilities for
configuration A and C are presented as function of baseline
length. The new data include the longest VLTI baseline so far and are
thus well suited to verify the size of the continuum region derived
from smaller baselines.   To facilitiate a direct comparison with D09, 
we compare the calibrated visibilities to a model of a Gaussian-shaped
emitting region with a full-width-at-half maximum (FWHM) of 0.98\,mas
and of 0.70\,mas respectively. This former value corresponds to the
stellar diameter derived by D09 based on the low spectral resolution
AMBER data, and baseline lengths extending up to 96 meters.   Future
observations at longer baselines, such as e.g. provided by CHARA
should allow us to distinguish between the various possible models for
the stellar photosphere such as a uniform disk vs. a Gaussian, as
applied here.  Note that the model visibilities in Fig.\,\ref{viscal}
do not reach unity at zero spacing but rather 94\%, because of the
continuum contribution by the over-resolved dust component within the
field of view of the UTs.  This fraction was determined from the field
of view analysis for the UTs presented in D09. This larger scale dust
emission has been modelled previously in e.g. Bl\"{o}cker et
al. (1999).

The new AMBER data are largely consistent with an emitting region of
0.98\,mas, although the measurement at the longest baseline hints at a
smaller diameter. A simple calculation shows that for a luminosity of
\IR of $\rm 25\,462~(d/kpc)^2~L_{\odot}$ (Bl\"{o}cker et al. 1999) and
a mid-A spectral type (Oudmaijer et al. 1996), a stellar diameter of
$\sim 0.70$\,mas can be expected (see Fig.\,\ref{viscal}). This value
is independent of the adopted distance. The substantial uncertainties
on the calibrated visibilities presented here render the reality of
this apparent discrepancy unsure, and more precise measurements and
longer baseline lengths are required for a better assessment. Moreover
we note that the calibrator star used with the FINITO observations is
somewhat brighter and therefore could introduce a bias towards lower
calibrated target visibilities. For this paper, we adopt a value for
the photospheric diameter at the {\it K} band of 0.98\,mas in order to
facilitate direct comparisons with Driebe et al. (2009). The continuum
visibilities for the new observations presented in this paper are
adopted to be the values of a Gaussian with a FWHM of 0.98\,mas  taking
into account the 6\% flux contribution by the over-resolved dust
emitting structure.

 Finally, we note that in medium spectral resolution the \Brg line
profile has a FWHM corresponding to the instrumental resolution, and
thus remains spectrally unresolved.  The high spectral resolution (HR)
observations (D09) resolve the \Brg emission line, which has a FWHM of
$\rm \sim 68\,km\,s^{-1}$, a line to continuum ratio of 2.105 at
line-centre, and an equivalent width (EW) of $-6.7\AA$. The MR
spectral setting has $R=1500$ or $\rm \sim 200\,km\,s^{-1}$ at the
\Brg transition, which clearly renders the line profile
unresolved. The MR data have the same EW, but as a consequence of the
lower resolution the line is smeared out and has a smaller
line-to-continuum ratio of 1.45. The fact that the lines are
spectrally unresolved will, in a completely analogous manner, also
lead to a smaller drop in visibility across a line. This in turn
results in derived size scales that are too small. In
Sect.~\ref{efspres} we will perform a consistency check between the
data with different spectral resolution before analyzing the combined
sets of interferometric AMBER data on \IR.

\subsection{X-Shooter spectroscopy}

IRC +10420 was observed on the 15$^{\rm th}$ of August 2009 (UT),
during the Science Verification observations with the X-Shooter
instrument. X-Shooter takes the spectrum 0.3 to 2.4 $\mu$m in one
shot, using three arms which cover the UV-blue, visible and
near-infrared parts of the spectrum respectively (D'Odorico S. et
al. 2006\nocite{2006SPIE.6269E..98D}). It is mounted on UT2 (Kueyen)
of the ESO-VLT telescopes in Mount Paranal, Chile.  The observations
were conducted in nodding mode, the nodding step was 6 arcsec.  The
slit width was 1.0 arcsec in the UV-Blue arm, and 0.4 arcsec in both
the visible and the near-infrared arms.   The data were reduced using
the X-Shooter pipeline provided by ESO \citep{modigliani}, and most of the spectrum and
the details of their observations and reduction will be presented
elsewhere.

Here we present data from the near-infrared spectrum covered by our
AMBER observations. Although the on-target spectra were of good
quality, the nodded, ``sky'' position spectra around Br$\gamma$ and
the Na {\sc i} doublet at 2.20 $\mu$m were particularly affected by
bad pixels, and a satisfactory correction for these proved
difficult. We therefore present the data from the first 8 on-target
spectra only. The total integration time for these was 8 seconds,
resulting in a signal-to-noise ratio in the continuum around
Br$\gamma$ of more than 100. This quality was matched by that of the
object HD 183143, a B-type supergiant only a few degrees away from IRC
+10420, and which was used as a telluric standard.  The data were
reduced by simply extracting the on-target spectra and subtracting the
background. Given that IRC +10420 is very bright ({\it K} $\sim$ 3.5),
the background is negligible and not having applied the normal
target-sky subtraction does not affect the data at all.

The individual spectra were co-added after correcting for the slight
sub-pixel wavelength shifts between them. Wavelength calibration was
performed by identifying several telluric absorption lines in the
2.15-2.2 $\mu$m region using the catalogue by Hinkle et
al. (1995\nocite{1995PASP..107.1042H}). The resulting wavelengths
should be accurate to less than a tenth of the spectral resolution
which was determined to be around 10,000 from measuring the 
FWHM of many narrow, unresolved, telluric absorption lines
in the spectrum.  The seeing as measured from the spatial profile of
the spectrum was 1.7 pixels, corresponding to 0.35 arcsec.

Finally, the spectrum of IRC +10420 was divided by that of the
telluric standard star HD 183143 to remove the telluric absorption
features. Its spectrum was first corrected for its underlying
Br$\gamma$ absorption line, which was removed using multiple Gaussian
fits to its clearly visible absorption profile.

\section{Results}
\label{results}

\subsection{Short description of the spectrum}

\begin{figure}
 \includegraphics[width=9cm]{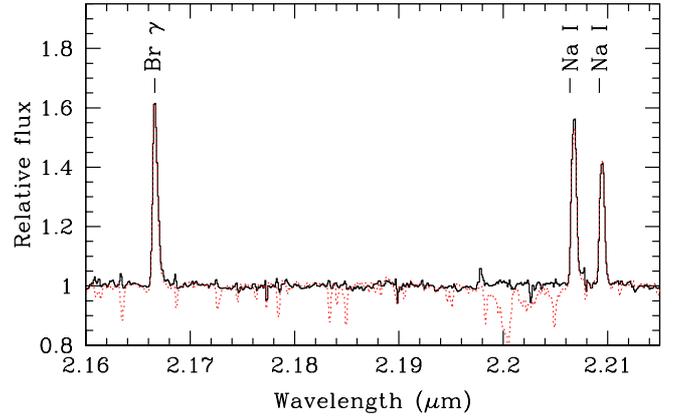}
\caption[]{Continuum rectified X-Shooter data around the wavelength
 range discussed here. The black line shows the final spectrum, the
 dotted line represents the spectrum uncorrected for telluric
 absorption.}
 \label{xsh}
\end{figure}

The X-Shooter spectrum in the wavelength range covered by our
AMBER-FINITO observations is shown in Fig.~\ref{xsh}. Both Br$\gamma$
and the Na {\sc i} doublet at 2.2$\mu$m are in emission.  The
line-to-continuum ratio of the Br$\gamma$ line is $\sim$1.6, which is
significantly smaller than the value of 2.1 reported by D09. This may
at first sight sound surprising as the spectral resolution of the
X-Shooter data is comparable to that of AMBER's high resolution
setting. However, contrary to AMBER which over-resolves the dust
continuum spatially, the X-Shooter spectrum includes the {\it K-}band
continuum excess emission. The line strength,  measured in terms
of EW as well as line-to-continuum ratio, is therefore smaller in the
X-Shooter data than in the AMBER data. As we will discuss later, the
same is the case for the Na {\sc i} doublet emission.

The line widths of the three emission lines are similar, of order 70
km s$^{-1}$, which, corrected for the spectral resolution, corresponds
to 63 km s$^{-1}$. This width is comparable to most optical emission
lines measured by Oudmaijer (1998\nocite{1998A&AS..129..541O}). The
central velocities of the emission lines are also similar, at 66 km
s$^{-1}$ (Local Standard of Rest, LSR) the lines are blue-shifted by
about 10 km s$^{-1}$ with respect to the systemic velocity of 77 km
s$^{-1}$ measured by Oudmaijer et al. (1996) from the CO (sub-)mm
rotational lines. The equivalent width of the Br$\gamma$ line is
$-$4.1$\rm \AA$.  This compares well with the measurement by Humphreys
et al. (2002\nocite{2002AJ....124.1026H}) who reported a Br$\gamma$ EW
of $-4~\AA$ in their 2000 spectrum.  The EW of the Na {\sc i} emission is
$-$3.7 and $-$2.9 $\rm \AA$ for both components respectively, giving a
total EW for the doublet of $-$6.6 $\rm \AA$. Typical errors on the EW
values are 5-10 per cent.

\subsection{The spectro-interferometry, visibilities and phases}

\begin{figure}
 \includegraphics[height=9cm,width=7cm,angle=90]{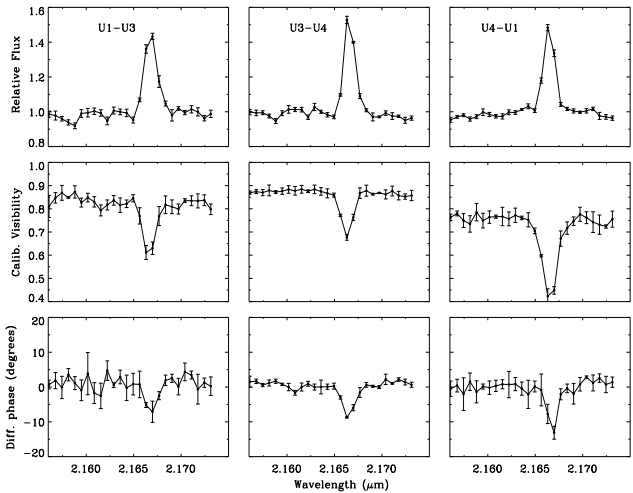}
 \caption[]{The configuration A observations. The top panels show the
normalized flux spectra, the middle panels show the visibility and the bottom
panels show the differential phase as a function of wavelength. }
 \label{vis1}
\end{figure}

\begin{figure}
 \includegraphics[height=9cm,width=7cm,angle=90]{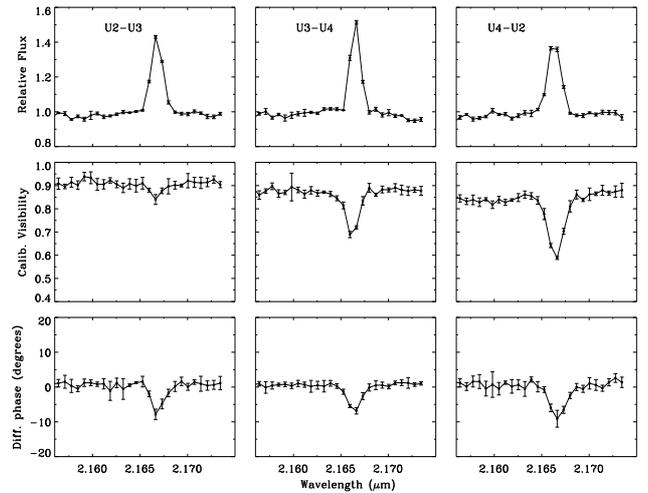}
 \caption[]{As the previous figure, but now for the configuration B observations.}
 \label{vis2}
\end{figure}

\begin{figure}
 \includegraphics[height=9cm,width=7cm,angle=90]{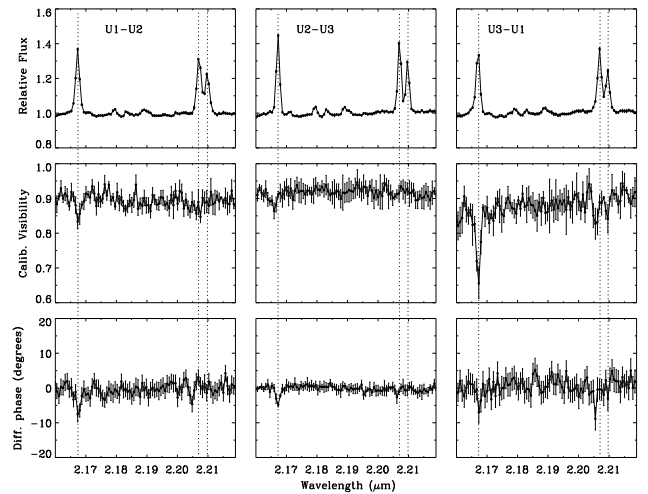}
 \caption[]{As the previous figure, but now for the configuration C
   observations. These observations were taken with FINITO, resulting
   in a a larger wavelength coverage than the other configurations. We
   show the range containing the \Brg and \ion{Na}{i} emission
   lines. Note that the drop in visibilities across line centre of the
   \ion{Na}{i} lines is consistently smaller than for \Brg at all
   baselines, suggesting the emission comes from a smaller region. The
   various faint spectral features are the result of imperfect
   cancellation of absorption features in the spectrum of the standard
   star (see Fig.~\ref{xsh}), this does not affect the visibility
   measurements.
\label{vis3}}
 
\end{figure}

The spectra, calibrated visibilities and differential phases for each
AMBER configuration are shown in Figs.\,\ref{vis1}, \ref{vis2}, and
\ref{vis3}. The panels of each figure show the reduced AMBER
observables for each baseline: the flux (top), visibility (middle) and
the differential phase (bottom) spectra.  No wavelength calibration
has been performed, and the spectra have simply been shifted in
wavelength in order to match the emission line with the Br$\gamma$
rest wavelength. The spectra have been divided by the interferometric
calibrator star in order to remove telluric absorption
components. What was evident in de Wit et al. (2008) and D09 is also
observed now at different baselines and position angles: a strong line
effect over \Brg in visibility and differential phase.  The drop in
visibility increases with baseline, i.e. the line is best resolved at
the longer baselines, implying that the line emitting region is small,
of order the resolution corresponding to the longest baselines. The
equivalent width of \Brg line is $-6.7\pm0.4~\rm \AA$ in the AMBER
spectrum taken in 2006 and the same is measured from the high
resolution AMBER spectrum taken in 2008. The flux spectra of the new
baselines reveal the same EW of $-6.7\pm0.4~\rm \AA$. These values of
the \Brg EW suggest that the line has not appreciably changed over the
past three years. This is in line with the EW of the H$\alpha$
emission line that varied little over that same time-span (Patel et
al. 2008\nocite{2008MNRAS.385..967P}).

A new result regarding the characteristic size-scales of the emission
line regions, and thanks to FINITO, involves the \ion{Na}{i}
transition for which we detect a faint but significant line effect in
the visibility on the longest baseline of configuration C
(Fig.\,\ref{vis3}). The small drop in visibilities at the \ion{Na}{i}
transition shows that this emission region is smaller than that of
\Brg but larger than the photosphere of the star.  A very rough
estimate for the characteristic size-scale of the \ion{Na}{i} emission
is obtained by calculating the line visibility using the flux-ratios
and compare this to what is expected for a Gaussian distribution at
the same spatial frequencies. Ignoring for the moment that the line is
spectrally unresolved and that the underlying star most likely
displays photospheric absorption, the line flux implies a size scale
of order two times that of the photosphere.

The Na {\sc i} line emission does not show a change in differential
phase across the lines unlike Br$\gamma$. Although there is an apparent
effect close to the Na {\sc i} lines, it does not coincide with the
wavelength of this doublet, and further inspection of the data reveals
that the SNR is too low to permit a robust measurement of a
differential phase at this transition.

\subsection{The field of view}

We reiterate that the AMBER line-to-continuum ratios and EWs are
higher than those measured from the normal long-slit, X-Shooter,
spectrum.  The reason for the discrepancy in EW between AMBER spectra
and long-slit spectroscopy is well understood; whereas the Br$\gamma$
line emission comes from well within the (limited) field of view of
the interferometric observations, the dust continuum emission falls
largely outside the field. This enhances the line flux with respect to
the remaining, mostly stellar, continuum.

For example, by adding 50\% of continuum (i.e. adding 0.5) to the
total continuum emission reduces the EW measured from the AMBER flux
spectrum from $-6.7\AA$ to $-4.5\AA$. For reference, adding only 40\%
continuum leads to a \Brg EW of $-5.0\AA$. In case of the \ion{Na}{i}
doublet, the X-Shooter EW measured over both lines is $-6.6\pm0.5\AA$.
The average AMBER spectrum however shows a \ion{Na}{i} EW of
$-9.1\pm0.4~\rm \AA$. Just like the \Brg emission, here we also need to add
around 40\% continuum emission to obtain the EW of the transition
taken with a long-slit spectrum. This leads us to conclude that the Na
{\sc i} emission almost entirely arises from within the
interferometric field-of-view of about 60\,mas, which is consistent
with the smaller visibility compared to Br$\gamma$ and indicates that
the emitting region falls well within that of the Br$\gamma$ emission.

\subsection{Consistency check between MR and HR visibilities}
\label{efspres}

We combine the sets of MR and HR visibilities for the \Brg transition
in order to verify the derived characteristic length scale of the
ionized wind by use of geometrical models. The large P.A. coverage
presented in this paper allows us to establish any deviations from a
centro-symmetric emitting region on milli-arcsecond scales, which is
relevant in light of previous claims (e.g. Davies et
al. 2007\nocite{2007ApJ...671.2059D}).  As noted earlier, the
Br$\gamma$ emission remains spectrally unresolved in the MR
observations, reducing the difference between line and continuum
visibilities.  Usage of spectrally unresolved visibilities results in
an underestimate of the true size of the line emitting region.  Here
we perform a consistency check between the high spectral resolution
and medium spectral resolution visibilities by applying a geometrical
model, originally proposed in D09.

The D09 geometrical model consists of three components: (1) an
over-resolved continuum emitting region, most of whose emission falls
outside the AMBER field-of-view of 60 mas when the UTs are employed;
(2) a star represented by a Gaussian with a FWHM of 0.98\,mas; and (3)
an ionized wind in the form of a ring with diameter of 4.18\,mas. D09
estimate that 6\% of the total continuum flux comes from  the dust
shell within the VLTI-AMBER field of view, the remainder of the
continuum flux in the AMBER spectrum is due to the star.

We should note that these values were adopted by D09 for the entire
wavelength ranges. However, the fractional flux contribution from the
dust continuum, like that of the stellar continuum, changes at the
\Brg line centre, where the line emission of course also contributes
to the total flux. The over-resolved component's contribution can then
only be 2.9\% of the total flux where the emission line peaks. 
Thus, when we correct for this, the ionized ring contributes 52.5\% at
line centre, and the star 44.6\% to the total flux, assuming for now
that there is no photospheric absorption. Inserting these corrected
flux contributions and the associated absolute calibration of the HR
visibilities (see D09) in the geometrical model, we find that the HR
visibilities are not well reproduced. This is especially true for the
longest baseline, where the relatively large contribution from the
small star pushes the geometrical model visibilities to higher values.
The discrepancy can be remedied by increasing the relative flux
contribution of the ionized wind to 58\%, which in turn implies that
\IR has an atmospheric absorption of 0.88, line-to-continuum
ratio. Some values are thus slightly different from those found in
D09, because of their oversimplified assumption of constant fractional
continuum-flux contributions over the whole wavelength range.

\begin{figure}
  \includegraphics[height=9cm,width=7cm,angle=90]{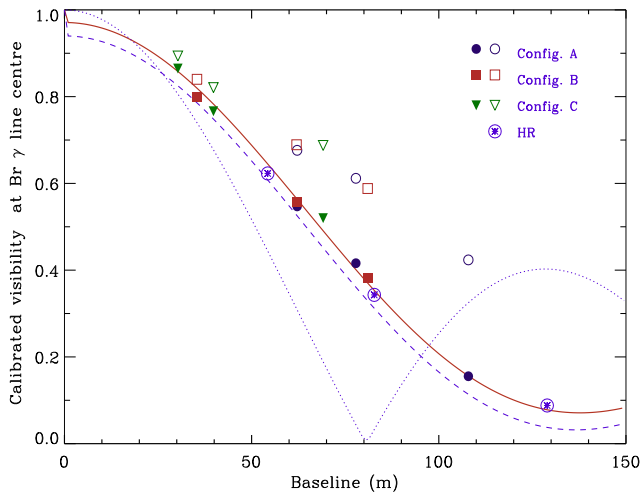}
 \caption[]{Observed and corrected visibilities at \Brg line
   centre. Open symbols are MR data uncorrected for spectral smearing,
   filled symbols are the same data points corrected for
   resolution. The three open circles with central dots are the high
   resolution observations.  These data are re-derived assuming a 2.9\%
   contribution by dust within the AMBER field of view at line centre.
   The corrected MR visibilities agree very well with the HR
   observations.  The full line represents visibilities from a model
   which includes a ring shaped ionised wind with diameter of
   4.18\,mas contributing 58\% to the total flux at line-centre (see
   text). 
The dashed line is the model from
   D09. Finally, to illustrate the the transition from first to second
   lobe at 80m, the dotted line is the the visibility function for a
   single ring with a diameter of 4.18\,mas.}
 \label{size}
\end{figure}

With the model values for the respective sizes in place, we can
calculate the visibilities for any baseline-length and any spectral
resolution. The resolved flux profile is approximated by a simple
Gaussian profile. Spectrally unresolved visibilities are calculated by
applying the flux ratios according to the convolved flux line profiles
(stellar photosphere and ionized wind) adopting a Gaussian kernel with
a FWHM equal to the spectral resolution under consideration.  Using
this approach, we can compute the multiplicative factor that corrects
the observed MR visibilities for spectral smearing. In this way we can
check for consistency between the two MR and HR data sets.  The
results are presented in Fig.\,\ref{size} for the total visibility at
line centre.  The open symbols represent the observed MR observations,
the filled symbols are the visibilities corrected for spectral
resolution. Also plotted are the three observed HR visibility points.
The corrected MR visibilities are consistent with the geometrical
model derived from the HR data and we find a size of 4.18\,mas for the
diameter of a ring-shaped ionised wind.  It may be clear how
significant the effect of spectral resolution in this case is. Based
on the MR data alone, one would derive a value for the diameter which
is about a factor of 1.5 too small (de Wit et al. 2008).
Figure\,\ref{size} also shows that the emitting region is very close
to centro-symmetric, at least at the line-centre of the \Brg
transition, because the {\it uv}-coverage extends in P.A. a range of
nearly 180\degr.

A third important conclusion that can be drawn from Fig.\,\ref{size}
is that the data for baselines longer than approximately 80m
probes the higher order lobes of the 4.18\,mas ring. This is illustrated
in Fig.\,\ref{size} by the dotted line, which represents the
analytical visibility function for a simple ring. This effect, as we
will detail in Sect.\,\ref{disc}, is key to understanding the
inversions of the differential phase in the HR data.

In conclusion, Fig.\,\ref{size} shows that (1) the spectrally
unresolved medium resolution visibilities are consistent with the
spectrally resolved high resolution observations, that (2) the
visbilities are consistent with an ionized wind that has a ring(-like)
morphology, and that (3) higher order visibility lobes are probed for
the ionized wind at baselines longer than 80m.

\section{Discussion}
\label{disc}
We have achieved a large increase in {\it uv}-coverage of AMBER
observations in the MR setting of the enigmatic star IRC
+10420. Interferometric studies so far have concentrated on the
ionized wind as traced by the \Brg emission. This emission remains
spectrally unresolved in the MR AMBER setting. Observations in the
high spectral resolution mode were presented for a single baseline in
D09 resolving the \Brg spectral profile. We will
discuss our new observational results in concert with the spectrally
resolved findings.

\subsection{Geometry of the \Brg line emission region, revisited}

The spectrally resolved \Brg emission line of IRC\,$+10\,420$ has
intriguing properties that the 2-D axisymmetric radiative transfer
code CMFGEN (Busche \& Hillier 2005\nocite{2005AJ....129..454B})
cannot reproduce (D09). Whereas the visibilities are reasonably
approximated, the model fails to reproduce the \Brg flux profile, its
amplitude, and the profile and amplitude of the phase signal. This led
D09 to conclude that the line emitting region deviates from spherical
symmetry which they accounted for by introducing an ad-hoc opacity
screen.  The main challenge with the (differential) phase signal is
its profile: the photocentres of the blue- and red-shifted part of the
wind are on the same side of the continuum. On the two longest VLTI
baselines of the HR observations, the differential phase inverts
giving the overall spectral phase profile an appearance similar to the
letter ``W'', with a stronger signal on the red-shifted part (see
Fig.\,\ref{hres}).

In order to understand this, we recall our analysis in the previous
section which supports the idea that at baselines exceeding 80m the
2nd lobe of the \Brg emitting ring visibility function is probed. A
change in the visibilities from one lobe to another is accompanied by
a 180$^{\rm o}$ phase flip. However, since not all emission at the
\Brg wavelength is resolved, the observed change will be smaller.  The
visibility profile on the 128.9m baseline (see Fig.\,\ref{hres})
displays a flat bottom and the phase discontinuities correspond
closely to the visibility discontinuities. These properties constitute
evidence for a transition into the 2nd lobe of the \Brg emitting ring.

\begin{figure}[t]
 \includegraphics[height=8cm,width=8cm]{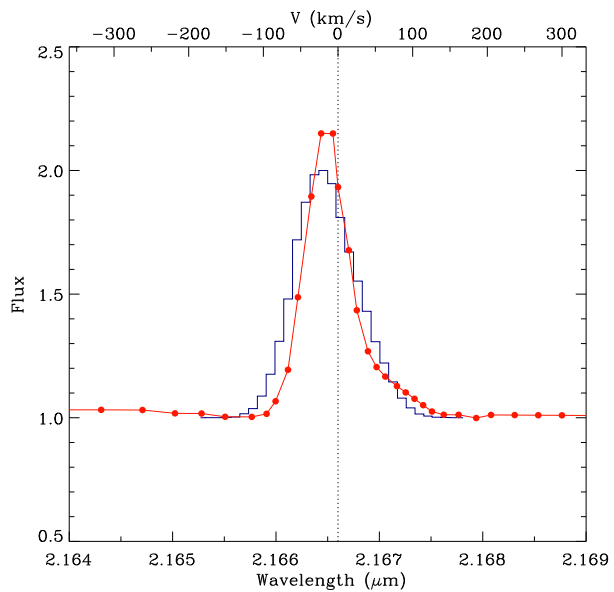}
 \caption[]{The observed AMBER HR flux profile is represented by the full line with small dots. The histogram 
shows the model spectrum which results from the geometric model with equal intensities for the \Brg emitting region.}
 \label{hresflux}
\end{figure}

\subsubsection{A geometric model}

Instead of using an advanced radiative transfer code, we attempt here
to find a simple geometry that would provide insight into the physics
of the ionized wind on milli-arcsecond size scales. Results from such
an exercise will constitute a good starting point for intensive
radiative transport modelling. An important observation is that the
peak flux of the Br$\gamma$ lines is blueshifted by 10 km s$^{-1}$
from the systemic velocity.  It has been suggested that this net
blueshift is because part of the redshifted wind is actually obscured
by the star (Oudmaijer et al. 1994).  In this context, it is worth
mentioning the multi-epoch HST observations by Tiffany et
al. (2010\nocite{2010AJ....140..339T}) from which the authors find
support for the idea that the \IR system is viewed close to pole-on.
The HR phase profiles (but also the flux profiles) preclude a \Brg
ring in Keplerian rotation, as this would result in different phase
shifts for the blue- and redshifted emission as for example
illustrated in Kraus et al. (2012)\nocite{2012ApJ...744...19K}.

As a first approach we assume that the wind is polar, consisting of
two spatially distinct yet similar components.  For simplicity reasons
and because the structure at line-centre is consistent with a ring
shape (Sect.\,\ref{efspres}), we assume a blue-shifted and a
red-shifted ring within the interferometer's field of view. The \Brg
emitting rings could make up the top and bottom part of an hour-glass
shaped wind geometry. Such a geometry is inspired by the photogenic
``Hourglass Nebula'' (the planetary nebula \object{MyCn 18}) which
shows evidence for multiple rings in H$\alpha$\, (see Dayal et
al. 2000\nocite{2000AJ....119..315D}). Finally, only a certain part of
the receding ring can be seen, because the material is partly obscured
by the stellar disk. This is possible, provided that the system is
viewed close to pole-on.

\begin{figure}
 \includegraphics[height=4.1cm,width=9.2cm]{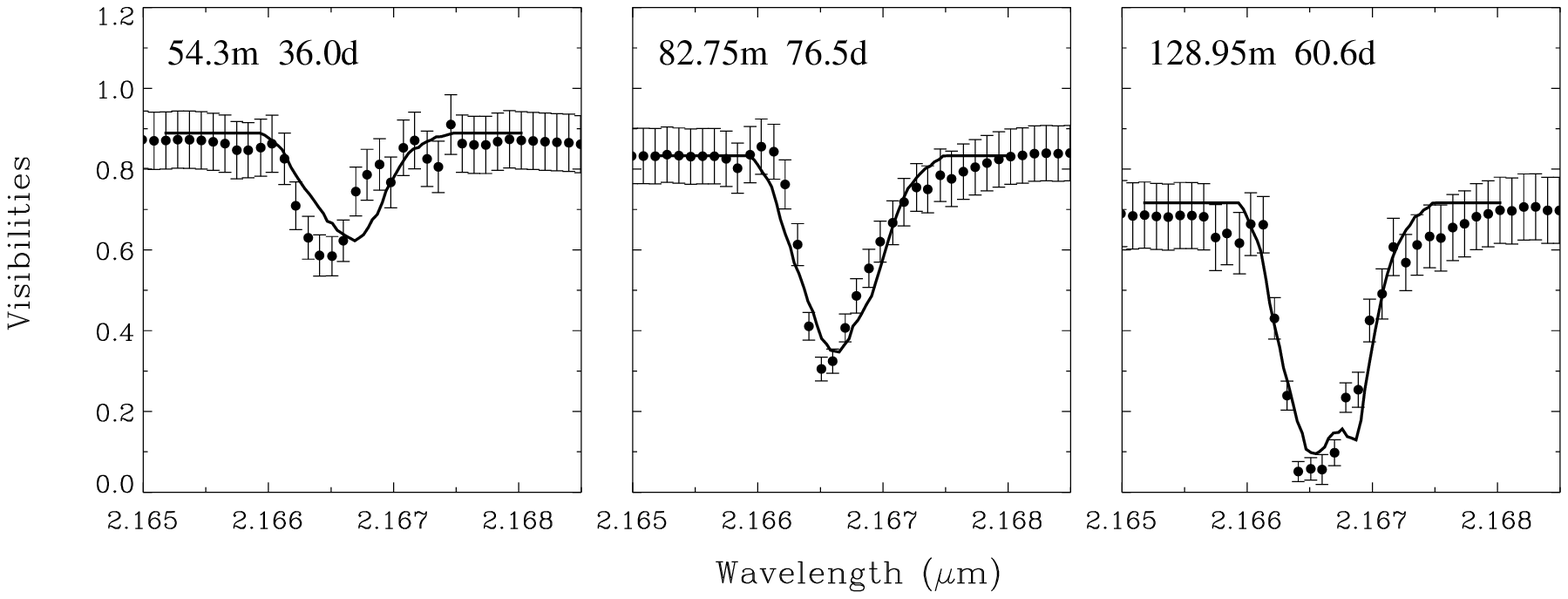}
 \includegraphics[height=4.1cm, width=9.2cm]{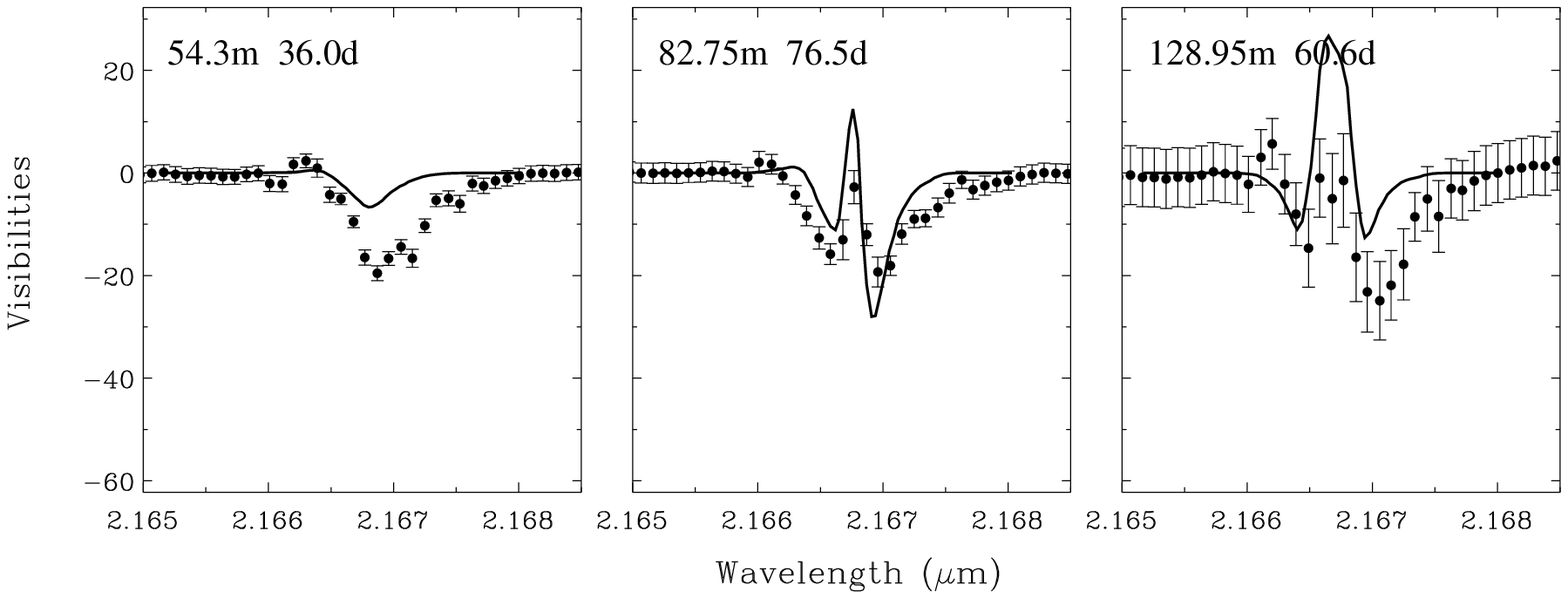}
 \caption[]{Computed (full line) and observed HR visibility and differential phase profiles of \Brg.}
 \label{hres}
\end{figure}

\begin{figure*}
\includegraphics[width=8cm,height=10cm]{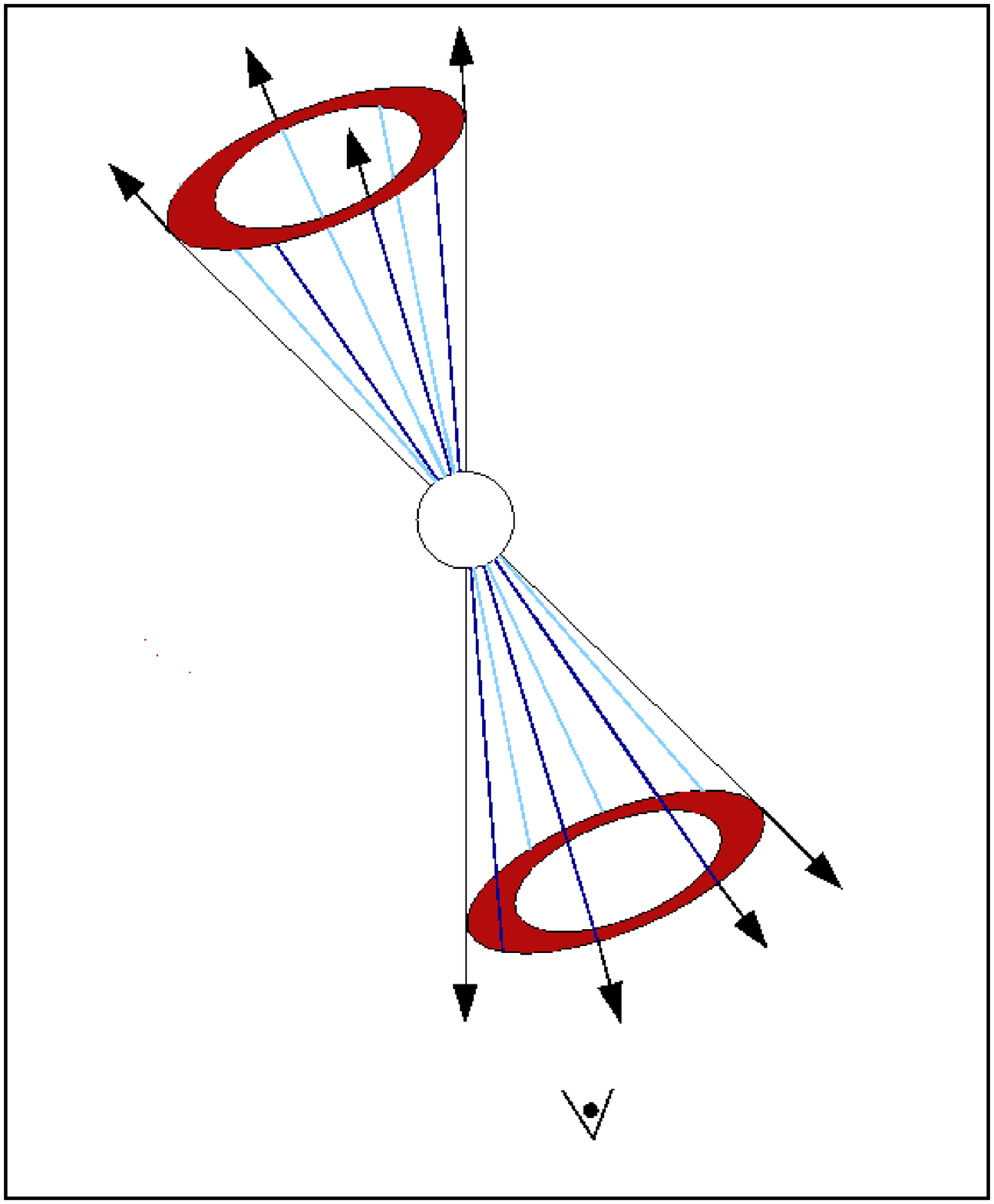}
\includegraphics[width=10cm,height=10cm]{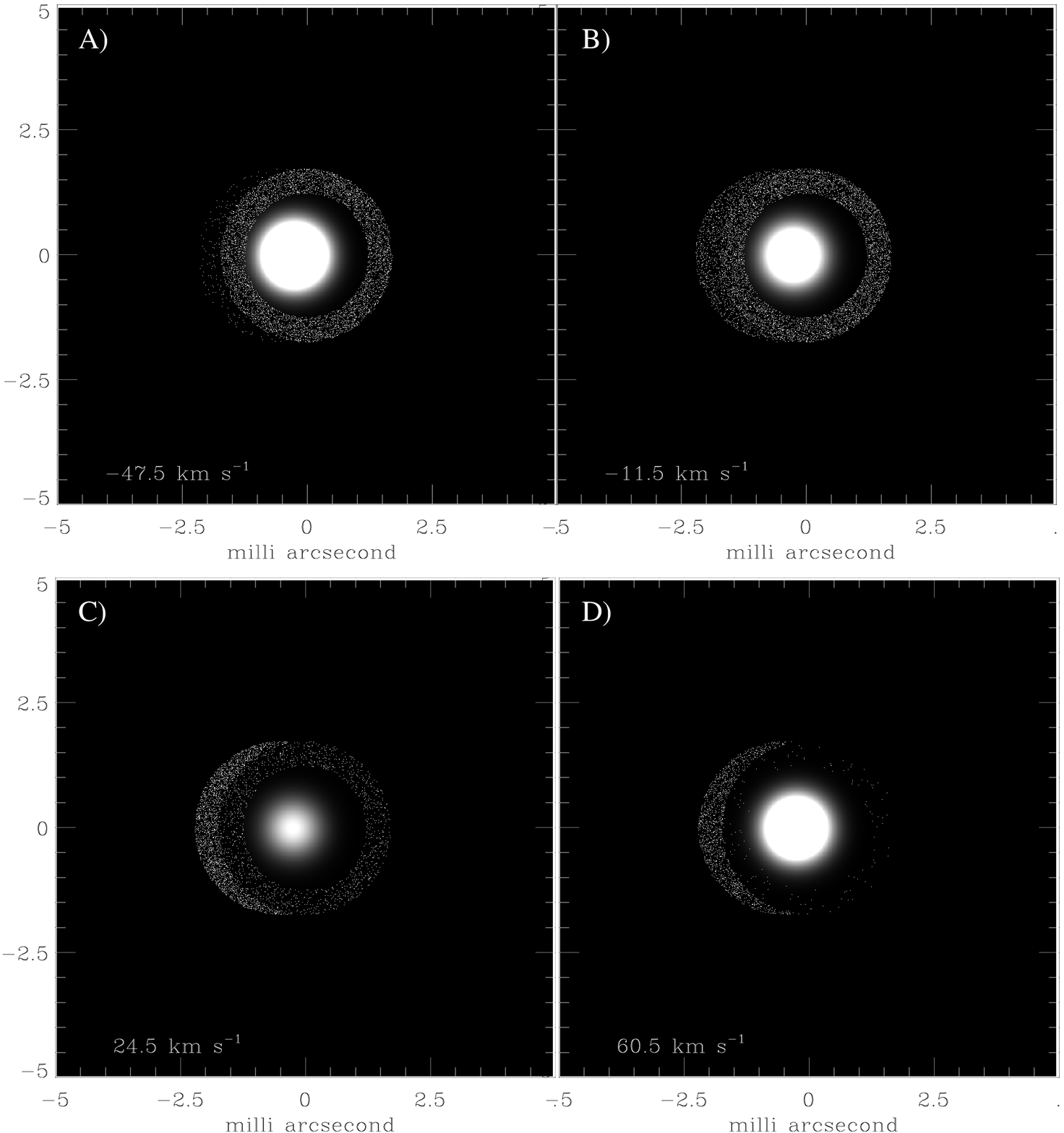}
 \caption[]{{\it Left:} 
Schematic representation of the model geometry.The observer is indicated by the symbol on the bottom of the graph. {\it Right:} Four
 model images at progressive velocity intervals. Consecutively, the
 images show A) the approaching ring; B) appearance of the unobscured
 part of the receding ring; C) disappearance of approaching ring; D)
 the receding ring with the right part obscured by the star. }
 \label{velim}
\end{figure*}

The model adopts a finite width ring with an inner and outer radius of
3.5\,mas and 4.5\,mas respectively The system needs to be inclined by
8\degr \, in order to have the star obscuring parts of the ionized
wind. The hourglass wind configuration is chosen to have an opening
angle of $2\theta=45\degr$. The combination of the system inclination
and the cone-opening angle is degenerate and neither are strongly
constrained: a wider cone opening angle requires a larger inclination
angle.  Indeed, we should perhaps stress that the model we construct
is only to be seen as a plausibility check and a start for further
studies.  The rings have an average outflow velocity of $\rm \pm
20\,km\,s^{-1}$ for the red and blue side respectively. The velocity
vector is parallel to the cone surface. In order to obtain higher
velocity material as observed in the flux profile, we apply a
multiplicative factor drawn from a Gaussian distribution with a FWHM
of 4. This results in velocity distributions with a FWHM of $\rm
40\,km\,s^{-1}$ for each ring.  If our model ring-elements at each
velocity interval contribute equally to the flux we obtain a \Brg
profile shown as a histogram in Fig.\,\ref{hresflux}. Our procedure
matches the observed flux profile by scaling the intensity of the
uniformly emitting ring-elements and thus introducing that the
ring-elements at some velocities are slightly brighter than others.

\begin{figure*}
 \includegraphics[height=9cm,width=9cm]{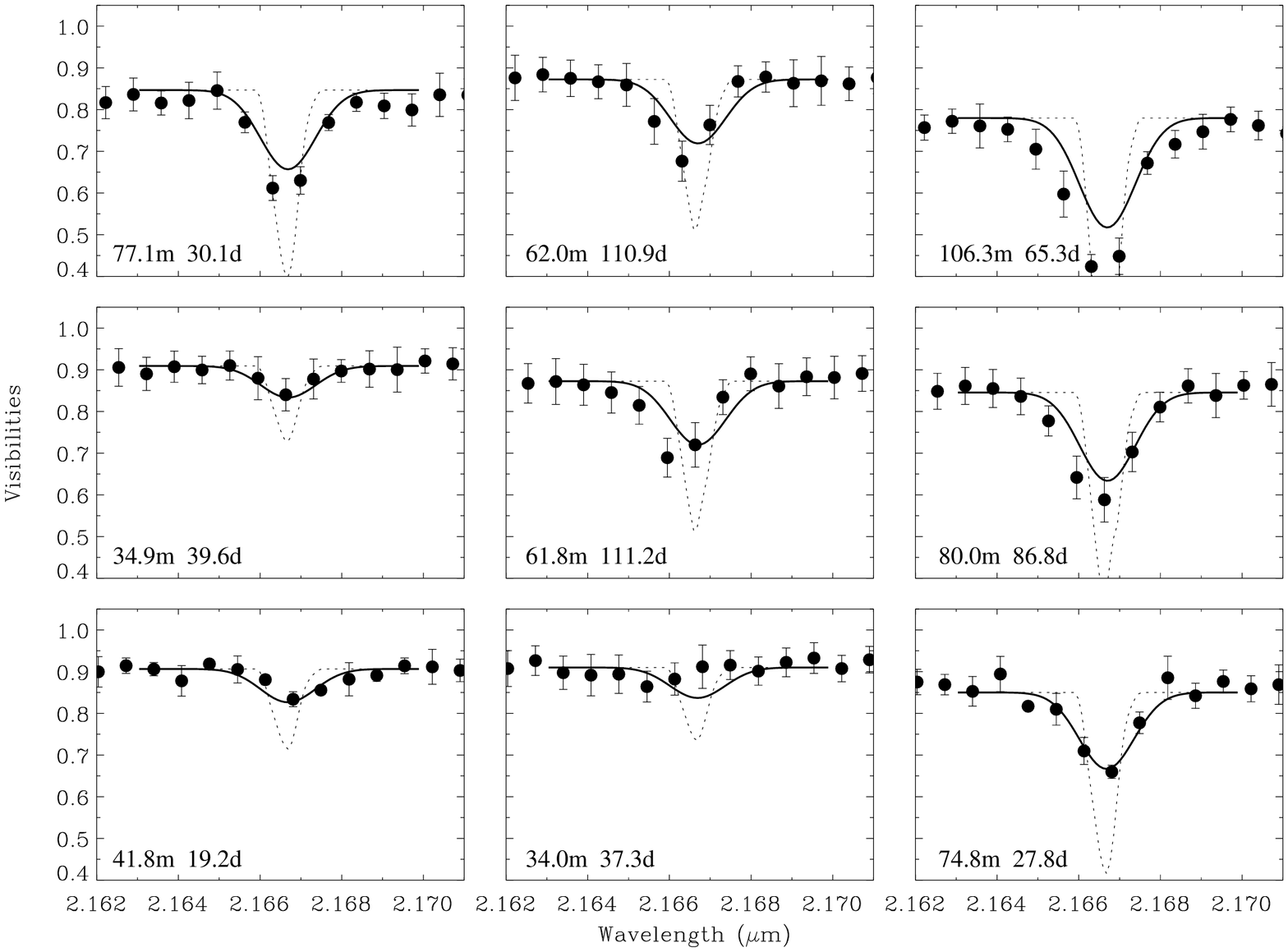}
 \includegraphics[height=9cm,width=9cm]{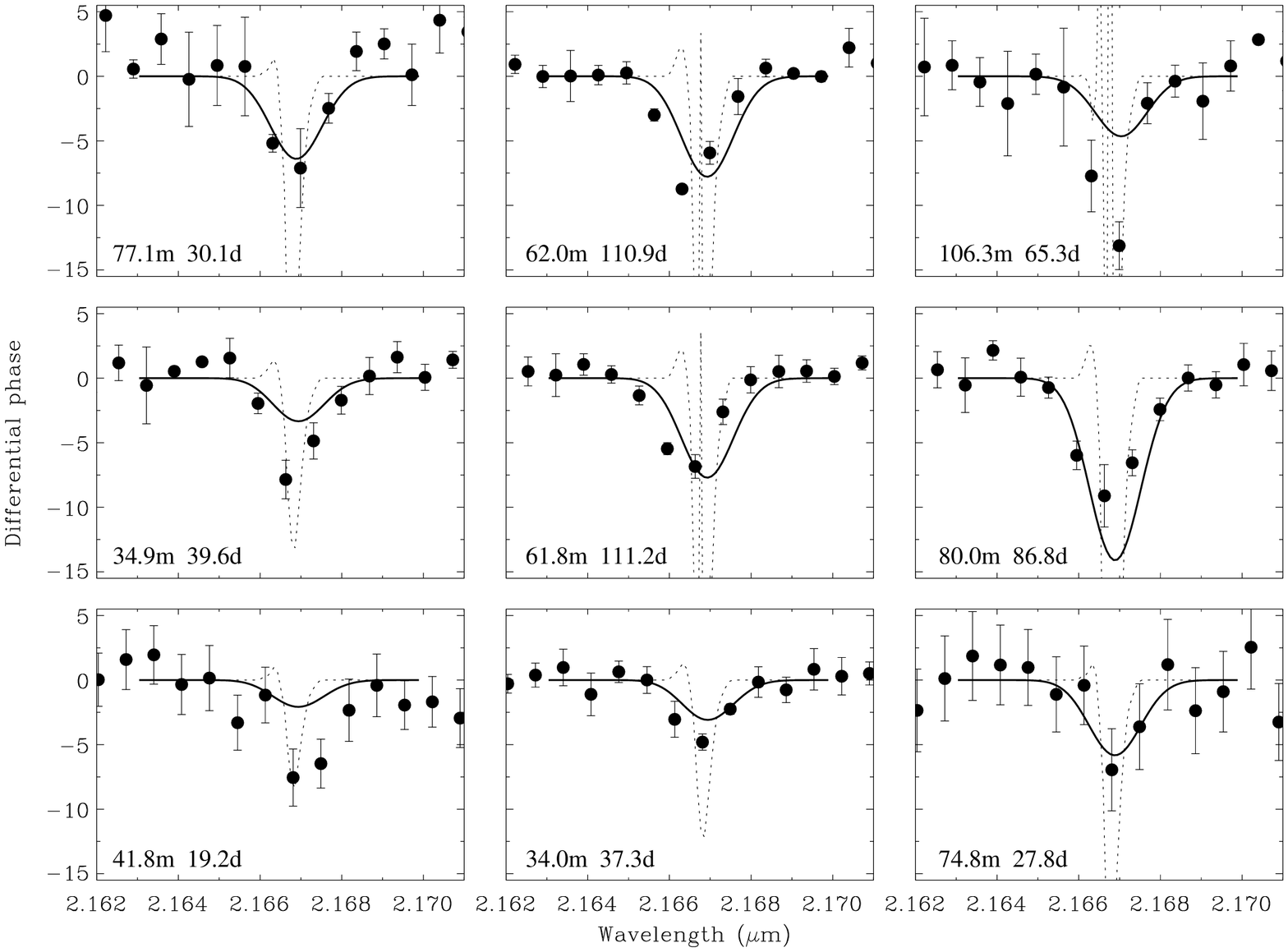}
 \caption[]{Measured and computed visibilities (left) and differential phases (right) of \Brg 
using the model presented in Sect. 4.1 for medium spectral resolution setting. Thick lines correspond to the model values at 
a spectral resolution of 1500. They are the result of a convolution a Gaussian with the dashed line, and the
later constitutes the spectrally resolved calculated profile.}
 \label{vismedres}
\end{figure*}

Using the fluxes thus obtained we calculate images at each velocity
interval. Fig.\,\ref{velim} displays the image on the sky for 4
velocity intervals, and it shows that the redshifted \Brg ring is only
partially observable.  From these images, we calculate the spectrally
resolved visibilities and phases which are presented in
Figs. \,\ref{hres}.  The visibilities show a decent match, and the
differential phases display a phase reversal because of the transition
into the second lobe of the ring's visibility function. By spectrally
degrading the model visibilities and phases to a spectral resolution
of 1500, we can compare the new nine AMBER observations in MR setting
to this simple model. These comparisons are shown in
Fig.\,\ref{vismedres}, where it is clear that again the visibilities
are nearly all well reproduced. Also the differential phases are in
general quite well reproduced, with some exceptions. Departures from
perfect ring structures in a dynamical environment can occur and could
be responsible for the three MR phases that do not fit the model
predictions.

We introduced a large macroturbulent velocity for the pragmatic reason
that we need to obtain a smooth \Brg line profile in the
model. However, this is not inconsistent with the qualitative
discussion of the physical state of the wind of IRC~+10420 by
Humphreys et al. (2002).  These authors invokes a picture of a dynamic
environment in which clumps move radially outward and inward.  Such an
environment is prone to shocks and this could provide the energy
dissipation that gives rise to the \Brg emission. It may well be that
the Br$\gamma$ emission is due to collisional excitation to level
$n$=2 followed by ionization and recombination in this dense
material. It would explain why a relatively cool object such as IRC
+10420 has such a prodigious hydrogen emission.

Our assumption of a ring-like structure is clearly an
oversimplification.  However, we note the similarity of the au-scale
environment with the clumpy structures in the larger scale ejecta
inferred from HST imaging (Tiffany et al. 2010).  We conclude with
summarizing the scenario in which a ring-like structure is the
morphological result of a dynamical outflow along an hour-glass like
cone nebula, giving rise to collisionally excited \Brg emission. The
model provides encouraging fits to both the observed visibilities and
differential phases obtained at many points in the {\it uv}-plane, and
should provide a good basis for more complex modelling of the
circumstellar environment of the object..

\subsection{The origin of the Na {\sc i} emission}

The Na {\sc i} 2.2 $\mu$m doublet in emission is a comparatively rare
phenomenon. It was first reported in the spectrum of IRC +10420 by
Thompson \& Boroson (1977\nocite{1977ApJ...216L..75T}), and has been
found in the spectra of other massive evolved stars, such as the
well-known Yellow Hypergiants HD 179821, HR 8752 and $\rho$ Cas
(Lambert et al. 1981\nocite{1981ApJ...248..638L}; Hrivnak, Kwok \&
Geballe 1994\nocite{1994ApJ...420..783H} Hanson et
al. 1996\nocite{1996ApJS..107..281H}), but also in Luminous Blue
Variables (LBVs) and B[e] stars (e.g.  Hamann \& Simon
1986\nocite{1986ApJ...311..909H}; McGregor et al.
1988\nocite{1988ApJ...324.1071M}; and Morris et
al. 1996\nocite{1996ApJ...470..597M}).  Several hypotheses have been
put forward to explain emission from a neutral atomic line towards hot
and cool stars. Here we investigate whether our spatially resolved
emission can constrain or rule out models.  We recall that the
spectro-interferometry indicates that all Na {\sc i} emission comes
from a region well within the dust condensation radius (70 mas). In
addition, we find that this atomic line emission originates from a
region smaller than the Br$\gamma$ emission.

The upper level of the Na {\sc i} doublet is at a much
higher excitation energy than normal for optical/near-infrared neutral
metallic lines.  Thompson \& Boroson (1977) proposed that the sodium lines
come from fluorescent emission occurring after the atoms are excited
to the upper state via pumping from the ground state by 3302 $\rm \AA$
photons.  To prevent the sodium from being ionized, shielding from
direct star light is necessary. Concentrating on the similar Mg {\sc
i} 1.5 and 1.7 $\mu$m emission lines in IRC +10420, Fix \& Cobb
(1987\nocite{1987ApJ...312..290F}) take this further and suggest the
neutral metallic lines are formed in a dense ``chromosphere'', located
just above the photosphere.

A different interpretation for the line forming mechanism can be found
in Hamann \& Simon (1986\nocite{1986ApJ...311..909H}).  These authors
associate the Na {\sc i} line emission of the - much hotter - B[e]
object MWC 349A with a cool, dense extended disk material and locate
the emission region 9-30 au from the B-type central star.  Similar
considerations led McGregor et al. (1988) to conclude that the Na {\sc
i} lines originate from the same region as the CO first overtone
bandhead emission, which typically arises from warm, dense
material. The density in this region would then be enhanced due to a
swept-up wind, and was proposed to be a few hundred au away from the
central stars.  Scoville et al. (1983\nocite{1983ApJ...275..201S})
observed the Becklin-Neugebauer object (2MASS05351411-0522227) in
OMC-1, a young, hot star which also has Na {\sc i} 2.2 $\mu$m
emission.  They locate the Na {\sc i} region outside the H {\sc ii}
region and propose that the Balmer bound-free continuum emission is
more likely to excite the atoms.

Our observations clearly rule out any scales larger than several 100s
of au, simply because of field-of-view considerations. However, scales
larger than 10s of au can also be excluded. The Na {\sc i} line
forming region is smaller than the Br$\gamma$ emission region for
which we derive a size scale of 4 milli-arcseconds or 4 stellar
radii. This finding lends support to the ``chromosphere'' scenario as
originally proposed by Thompson \& Boroson (1977) and Fix \& Cobb
(1987). The fact that the lines are blue-shifted with respect to the
systemic velocity is consistent with this idea. It is inevitable that
part of the red-shifted emission from a volume barely larger than the
star will be blocked from view. The rather smooth and symmetric
appearance of the lines probably indicates a large macro-turbulent
velocity in this wind.

The question that remains of course is whether we deal with a proper
``chromosphere'', as neither IRC +10420, nor the other Yellow
Hypergiants HD 179821, $\rho$ Cas and HR 8752, who also have Na {\sc
i} emission, show a temperature inversion indicative of a
chromosphere. These objects are the spectroscopic twins of IRC +10420,
while HD 179821 is, next to IRC +10420, the only other object to be
proposed in the post-Red Supergiant phase (see the review by Oudmaijer
et al. 2009\nocite{2009ASPC..412...17O}). As these authors point out,
these two objects could very well undergo strong mass loss, where the
wind itself is optically thick and forms a pseudo-photosphere hiding
the star itself from view (see also Smith, Vink \& de Koter
2004\nocite{2004ApJ...615..475S}). The optically thick wind shields
the sodium from ionization by direct starlight, and produces the
conditions to get Na {\sc i} emission, as observed. This is consistent
with the fact that the line forming region is smaller than
Br$\gamma$. Indeed, the presence of an optically thick
pseudo-photosphere could also explain why the star is observed to be
larger in the {\it K} band continuum than would be expected for its
spectral type (0.98 vs 0.70 mas - see Sec.2).

Is the inferred presence of a pseudo-photosphere consistent with the
properties expected for massive evolved stars?  Yellow Hypergiants are
hypothesised to evolve from the red to the blue, crossing the
so-called Yellow Void, a hitherto fairly empty region in the HR
diagram (de Jager \& Nieuwenhuijzen
1997\nocite{1997MNRAS.290L..50D}). Some of them have been observed to
undergo strong mass-loss events, creating a pseudo-photosphere making
them appear to cross this Yellow Void again, but now back to the red
(e.g. Humphreys et al. 2006\nocite{2006AJ....131.2105H}; Lobel et
al. 2003\nocite{2003ApJ...583..923L}). Such events have been referred
to as ``bouncing'' (e.g. Stothers and Chin
2001\nocite{2001ApJ...560..934S}) against the Yellow Void, although it
would be more correct if referred to as bouncing against a ``Yellow
Wall'' or rather a ``White Wall'' given the temperatures involved
($\sim$8000-10,000 K).  It may also be useful to note that the Na {\sc
i} emission was present in the spectrum of IRC +10420 at a time when
there was no Br$\gamma$ emission. This was discovered only much later
by Oudmaijer et al. (1994). Therefore the star may have been
surrounded by a pseudo-photosphere much longer, while the increase in
temperature of the underlying star eventually gave rise to the
observed hydrogen recombination line emission. IRC +10420 has been
photometrically stable for about a decade now (see e.g. Patel et
al.2008), suggesting it has not changed its temperature and it
therefore seems to be leaning against this White Wall.  The presence
of a pseudo-photosphere and associated mass loss, may indicate that we
could witness a strong eruption soon.

A question that arises now is whether the pseudo-photosphere scenario
is also applicable to the LBVs and B[e] objects. LBVs are known to
undergo strong mass loss episodes and, when at their coolest,
hypothesised to be surrounded by pseudo-photospheres (see the
discussion by Vink 2009\nocite{2009arXiv0905.3338V}). Na {\sc i}
emission is absolutely consistent with this and may even be used to
infer the presence pseudo-photosphere in the absence of other
diagnostics.  Moving to higher temperatures, an intriguing, recent,
AMBER result is that Na {\sc i} 2.2 $\mu$m emission from the B[e]
object CPD -57$^{\rm o}$2874 is {\it unresolved}. This is accompanied
by a spatially resolved Br$\gamma$ emission line region which is
elongated with a size of 7$\times$13 au (Domiciano de Souza et
al. 2007\nocite{2007A&A...464...81D}). As also observed for IRC
+10420, the Na {\sc i} emission originates much closer to the 20\,000
K star than the hydrogen recombination line emission.  It would be
tempting to assume that the star exhibits a pseudo-photosphere too,
however the spectral energy distribution indicates a hot
photosphere. As the Na definitely needs shielding from the stellar
radiation, it is plausible that in this case the line originates from
a dense disk instead, consistent with the elongation observed in
Br$\gamma$. It will be interesting to see how a larger baseline
coverage will be able to distinguish between such possibilities.

\section{Conclusions}

In this paper we have presented new spectro-interferometric AMBER and
spectroscopic X-Shooter observations of the post-Red Supergiant IRC
+10420.  The medium spectral resolution AMBER observations provide
about 180 degree coverage of the object in {\it uv-}space. The data
spatially resolve the \Brg emission line and, for the first time the
Na {\sc i} 2.2 $\mu$m doublet emission.

Our main conclusions are the following. We re-confirm that the ionized
wind appears as a ring-like structure on the sky. It has a diameter of
approximately 4 milli-arcsecond or $\sim 1$\,au. In order to explain
the unusual differential phases, we consider a geometric model in
which the ring structure traced by the interferometry are actually two
rings on either side of the star as part of an hourglass nebula seen
close to pole-on. The model reproduces the spectrally dispersed
visibilities and phases very well.

The direct implication is that \IR exhibits a polar wind. Whether the
wind in the equatorial region has different properties (density,
terminal velocity) requires further investigation. The observed
velocities in \Brg can be understood in terms of small pockets or
blobs with velocities of $~40\,\rm km\,s^{-1}$, being ejected along
the hourglass cone surface.

In contrast, the Na {\sc i} 2.2 $\mu$m doublet, a comparatively rare
emission line, is found to originate inside the \Brg emission region
and just outside the continuum.  It is very unusual to find emission
from a neutral metallic line coming from a region that is closer to
the star than the hydrogen recombination line emission. This leads us
to speculate that the hydrogen is not ionized directly from the ground
level, but instead from the $n=2$ collisionally excited level. This
requires less energetic photons and could explain why we would see
such prodigious emission from  a comparatively cool star, while it
is consistent with the idea of high densities in the circumstellar
material.

It is very telling that despite its rarity all Yellow Hypergiants with
published near-infrared spectra display this doublet in emission. If
we are to assume that the size scales found for IRC +10420 are typical
for the other ones, they all may exhibit an extended, expanding,
pseudo-photosphere.  The combination of the atomic properties
requiring this line to be pumped by UV-blue photons and the observed,
small, size scale of the line emitting region allows us to conclude
that all Yellow Hypergiants have a "pseudo-photosphere" yielding
support to the notion of the bouncing against the Yellow Void/White
Wall hypothesis.

\begin{acknowledgements}

It is a pleasure to thank Antoine M\'erand for his advices on data
reduction, and Thomas Driebe for providing us with the high spectral
resolution AMBER data. WdW would like to thank the hospitality of
G. Weigelt and the Bonn IR interferometry group where this paper was
finalized.

\end{acknowledgements}

\bibliographystyle{aa}
\bibliography{irc.bib}

\end{document}